\documentclass[showpacs, twocolumn,prl ]{revtex4}

\usepackage{graphicx}
\newcommand{\be}{\begin{equation}}
\newcommand{\ee}{\end{equation}}

\begin{document}

\title{Separation of two regimes in a disordered boson Hubbard Model} 

\author{Ji-Woo Lee$^1$ and Min-Chul Cha$^2$} 
\affiliation {$^1$Department of Physics, Duke University, 
Box 90305, Durham, North Carolina 27708-0305, USA \\
$^{2}$Department of Applied Physics, Hanyang University, Ansan, 
Kyunggi-do 426-791, Korea}

\begin{abstract}
We study the insulator-to-superfluid transition in a two-dimensional
disordered boson Hubbard model at zero temperature for intermediate
strength of disorder at commensurate density.
Via Monte Carlo calculations of the correlation functions
in the integer current representation of the model,
we obtain the dynamical critical exponent $z=1.5 \pm 0.1$,
supporting the multi-critical behavior separating the strong and
weak disorder regimes.
Investigating the density profile, we suggest that the density fluctuations
due to the particle-hole excitations drive the transition in the weak
disorder regime.
\end{abstract}
\maketitle

Localization in strongly correlated systems has attracted attentions as
an important problem in condensed matter physics.
The interplay of disorder and interaction in these systems still
remains not fully understood.
Among those models adopted to investigate the role of disorder,
the boson Hubbard model is one of the most convenient models 
partly because the order parameter is well defined.
The insulator-to-superfluid transitions of this model\cite{Fisher89}
in low dimensions due to both interaction and disorder
have been studied intensively in the past decade
\cite{Scalettar91,Krauth91b,Singh92,Runge92,Wallin94,Pazmandi98,
Zhang95,Amico98,Herbut00}.
This problem also has many experimental counterparts such as
$^4$He in porous media\cite{Crowell97} and 
cold atoms in optical lattices\cite{Greiner02,Zoller03}.
Also the superconducting transitions at zero temperature
in thin films\cite{Goldman98} are believed to belong to the universality 
classes represented by the boson Hubbard model.
The phase transitions in small Josephson-junction arrays\cite{Fazio01}
and granular superconductors\cite{Gerber97}
have been described by similar models.

The boson Hubbard model can be written as
\be
H_{bH} = \frac{U}{2} \sum_{\bf r} (n_{\bf r} - \mu_{\bf r} )^2  
- t \sum_{\langle {\bf r}, {\bf r'} \rangle}
(b_{\bf r}^\dag b_{\bf r'} + b_{\bf r} b_{\bf r'}^\dag ),
\label{eq:bH}
\ee
where $b_{\bf r} (b_{\bf r}^\dag)$ is the boson destruction (creation)
operator at site ${\bf r}$,
$n_{\bf r} \equiv b_{\bf r}^\dag b_{\bf r}$ is the boson number operator,
and $U$ and $t$ are the on-site repulsion energy and
the hopping matrix element, respectively.
Here we assume that disorder is given in the local chemical potential
so that ${\bar \mu}-\Delta < \mu_{\bf r} <{\bar \mu}+\Delta$,
where ${\bar \mu}$ is the average chemical potential which controls
the total density of bosons.
The parameter $\Delta$ characterizes the strength of disorder.
Different realizations of disorder in random magnitude\cite{Prokofev03} or
random phase\cite{Phillips03} of the hopping matrix $t$ have been discussed
recently.

In pure case ($\Delta=0$), the insulating phase of the model
has a finite Mott energy gap for excitations,
and the density fluctuations are suppressed.
The Mott gap vanishes at the transition.
The relevant density fluctuations driving the phase transition, as
$t$ increases, for the commensurate density (${\bar \mu}=0$)
are the particle-hole excitations.
This nature of the transition results in the dynamical critical exponent $z=1$.
For an incommensurate density (${\bar \mu} \neq 0$), however,
either single particle or hole excitations are favored
since the chemical potential breaks the particle-hole symmetry globally.
In this case, we have $z=2$\cite{Fisher89}.

In the presence of disorder ($\Delta \neq 0$), the situation is subtle.
Since non-zero $\mu_{\bf r}$ breaks the particle-hole symmetry locally,
by adopting a single-particle picture, it has been argued
that an arbitrarily weak random potential will localize a single-particle
excitation as soon as the Mott gap vanishes.
In other words, near the transition where the Mott gap is smaller than $\Delta$,
it is always possible to find an arbitrarily large region\cite{Freericks96}
in which the chemical potential is uniformly shifted 
to make the single-particle excitation gap vanish.
Therefore, there exists an insulating phase, called Bose glass (BG) phase,
which has finite regions with vanishing gap.
Phase transition occurs when this region grows infinitely.
A hallmark of this transition is the fact that $z=d$
($d$ is the dimensionality of the system),
since the BG phase is compressible\cite{Fisher89}.

This scenario of the BG-to-SF transition governed by growing zero-gap regions
is certainly convincing for large $\bar \mu$ or large $\Delta$
since particles or holes cause the density fluctuations separately.
However, near ${\bar \mu}=0$ for weak disorder, there is a possibility that
the particle-hole excitations still survive to drive the transition
possiblely due to the existence of the so-called statistical
particle-hole symmetry\cite{Weichman96}, i.e., that this symmetry is restored
when a random potential is averaged on the scales of
diverging correlation length.
Therefore, it is an interesting question to check whether the particle-hole
excitations govern the transition in the presence of weak disorder
to change the nature of the transition from that of the BG-to-SF transition.

In the quantum rotor model equivalent to the boson Hubbard model
in the limit of large density,
a direct transition from the Mott insulator(MI) to SF transition
was reported\cite{Kisker97} 
for a weak disorder $\Delta=0.2$ at ${\bar \mu}=0$ with $z=1$,
instead of the BG-to-SF transition.
Subsequently some numerical evidences were provided, which supports
that the direct MI-to-SF transition occurs even for non-zero small ${\bar \mu}$
\cite{Park99}.
Further Monte Carlo studies\cite{Lee01} of the same model suggested that
the nature of the transition is indeed divided by a multi-critical line into
the strong disorder regime where the BG-to-SF transition
occurs and the weak disorder regime where the direct MI-to-SF transition occurs.
As an evidence for the existence of the multi-critical line which defines
the critical strength of the disorder, $\Delta_c ({\bar \mu})$,
the new value $z=1.35 \pm 0.05$ was found on the line
through the finite-size scaling behavior of the superfluid stiffness.
Also the possibility that weak disorder is screened out by the proliferation
of the particle-hole excitations near the transition around ${\bar \mu}=0$
was raised.

One of the key quantities closely related with the nature of the transition
is the dynamical critical exponent, $z$.
In this work, we calculate $z$ directly from the correlation functions
through Monte Carlo simulations.
The motivation is that the numerical value of $z$ obtained in this way
is insensitive to some scaling parameters such as the aspect ratio and
$K$ as far as $K \approx K_c$, in contrast to the finite-size scaling
behavior which is in general very sensitive to those parameters.
Therefore, it is another method to obtain the critical exponents
even though its accuracy would be poorer than
that of a finite-size scaling method.
We also investigate the density profile to see how the density fluctuations
are raised.

In the limit of large density, the boson Hubbard model can be reduced to
a quantum rotor model
\be
H_{rotor} = \frac{U}{2}\sum_{\bf r}
(\frac{1}{i}\frac{\partial}{\partial \theta_{\bf r}} - \mu_{\bf r})^2
-J \sum_{\langle {\bf r}, {\bf r'} \rangle }
\cos ({\theta}_{\bf r} - {\theta}_{\bf r'} ),
\label{eq:rotor}
\ee
where $J= 2 n_0 t$ with the average number of bosons per site $n_0$,
and $\theta_{\bf r}$ is the phase angle of the local order parameter.
This model has the global $U(1)$ symmetry ({\it i.e.\/} the Hamiltonian is
invariant under the transformation $\theta_{\bf r} \to \theta_{\bf r} + \phi$
for a constant $\phi$).
It implies that the ground state of this Hamiltonian has two phases: one
with this symmetry, which is an insulating phase, and the other, the
superfluid (SF) phase, in which the symmetry is spontaneously broken.

This transition has been studied in an equivalent classical model\cite{Wallin94}
whose partition function, represented by integer current vectors ${\bf J}$,
is given by
\be
Z(K) = \sum_{\bf \{J \}}^{\nabla \cdot {\bf J} = 0}
\exp \Big({ - \frac{1}{2K} \sum_{{\bf r},\tau}
{J^x_{\bf x}}^2 + {J^y_{\bf x}}^2
+(J^{\tau}_{\bf x} - \mu_{\bf r} )^2 } \Big),
\label{eq:classical}
\ee
where ${\bf x} = ({\bf r}, \tau)$ is $(d+1)$ dimensional lattice point,
and the expression, 
${\nabla \cdot {\bf J} = 0}$,
denotes the current conservation condition.
Here $K \sim \sqrt{t/U}$ is the parameter controlling the quantum fluctuations.

The correlation function, in terms of $\theta$ in Equation(\ref{eq:rotor}),
is defined as 
\be
C({\bf r}_1, {\bf r}_2, \tau_1, \tau_2) = [ \langle \exp \{ i{\theta}({\bf r}_1, \tau_1) - i {\theta} ({\bf r}_2, \tau_2 ) \} \rangle ]_{av},
\ee
where $\langle \cdots \rangle$ denotes the ensemble average
and $[\cdots]_{av}$ denotes the average over different disorder complexions
with the same strength.
In terms of the current variables, it can be rewritten as
\be
C_x(r)= [ \langle e^{-(1/K) \sum_{\rm path} (J^x_{\bf x}  + 1/2 )} \rangle ]_{av}
\ee
for the equal-time correlation functions, and
\be
C_\tau^{\pm} (\tau) = [ \langle e^{\mp (1/K) \sum_{\rm path }
(J^{\tau}_{\bf x} \pm 1/2 - \mu_{\bf r})}\rangle  ]_{av}.
\ee
for the zero-distance correlation functions.
$C_\tau^{\pm} (\tau)$ are the correlation functions for $\tau >0$ or $\tau <0$.
When ${\bar\mu}=0$, we expect $C_\tau^{\pm} (\tau)=C_\tau (\tau)$,
implying the global particle-hole symmetry.
Here a path is chosen which connects two points separated
in distance $r$ in a spatial direction or $\tau$ in the temporal direction.
Recently a new algorithm was developed\cite{Alet03a,Alet03b} for calculating
these correlation functions very efficiently.
By adopting this algorithm,
the correlation functions are calculated via Monte Carlo simulations in
(2+1) dimensional lattices near $K_c \approx 0.292$
when $\bar\mu=0$ and $\Delta=0.4$.
This is the critical strength of disorder within the error range.
The critical point $K_c$ is determined from
the finite-size scaling of the superfluid stiffness.
Typically the average is taken over 1,000 - 2,000 disorder realizations.
For each disorder configuration, we construct
2,000 worms and 40,000 worms
for the smallest system ($8\times 8\times 17$) and
the largest system ($24\times 24\times 73$)
in our simulations, respectively.

\begin{figure}
\includegraphics[width=7.0cm]{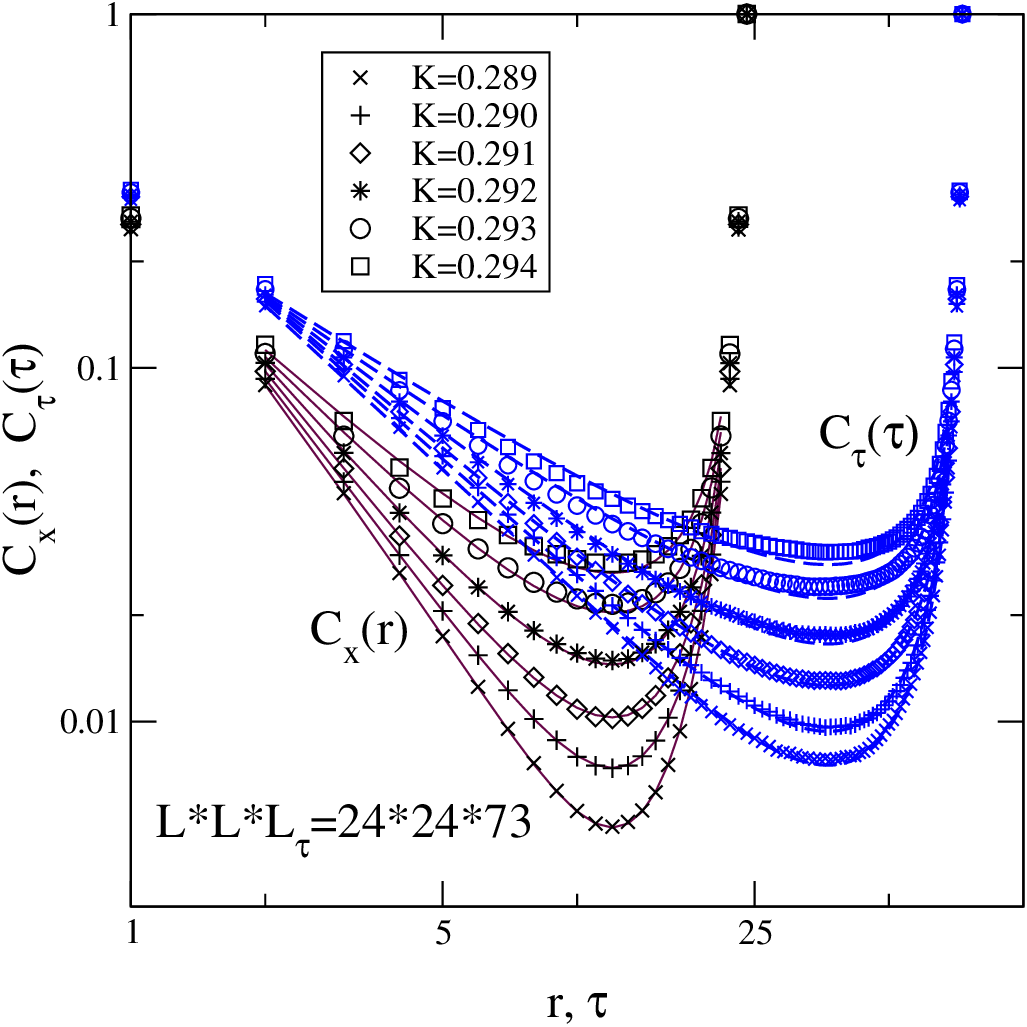}
\caption{
The correlation functions on $L \times L \times L_\tau=
24\times 24\times 73$ lattice near $K=0.292$.
By fitting the data, we find that $y_x=1.43$ and $y_\tau=0.97$ at $K=0.292$.
}
\label{fig:fig1}
\end{figure}

\begin{figure}
\includegraphics[width=7.0cm]{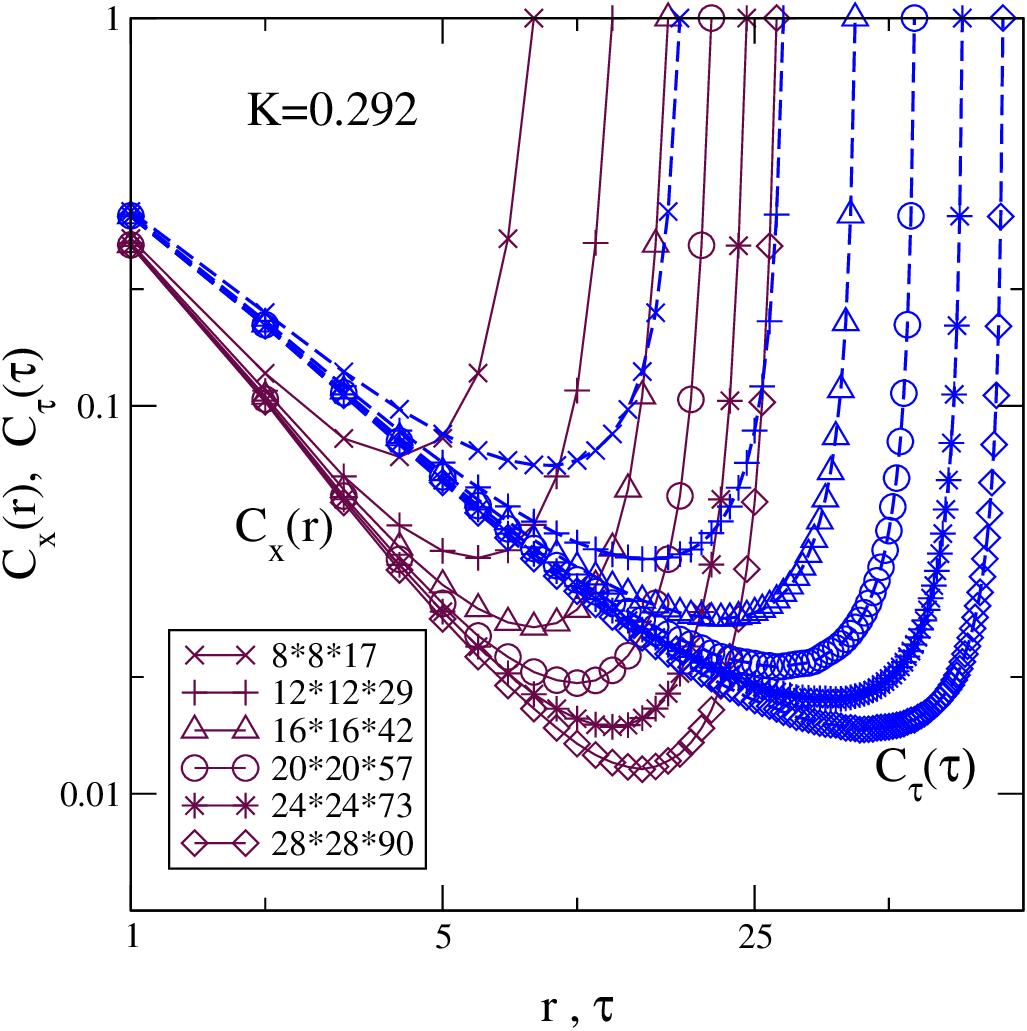}
\caption{
The correlation functions for $K=0.292$ in different sizes.
}
\label{fig:fig2}
\end{figure}

The results are shown in Figures \ref{fig:fig1} and \ref{fig:fig2}
in different sizes and for different $K$'s.
In order to extract the critical exponents, we fit the data on
scaling curves.
The scaling form for the correlation function is given by
\be
C(r, \tau) = (r^2+a\tau^{2/z})^{-(d+z-2 + \eta)/2} f( r/\xi , \tau/ \xi^z ),
\ee
where $f$ is a scaling function and $a$ is a non-universal constant.
At the critical point ($\xi \to \infty$),
the correlation functions in finite size systems will have forms
\be
C_x(r) = c\Big[ {e^{-r/\xi_x}\over r^{y_x}}
+ {e^{-(L-r)/\xi_x}\over (L- r)^{y_x}} \Big]
\ee
and 
\be
C_\tau(\tau) = c'\Big[ {e^{-\tau/\xi_\tau}\over \tau^{y_\tau}}
+ {e^{-(L_\tau-\tau)/\xi_\tau}\over (L_\tau- \tau)^{y_\tau}} \Big],
\ee
where $c$ and $c'$ are some constants, 
$\xi_x$ and $\xi_\tau$ are parameters characterizing the correlation lengths
in $x$ and $\tau$ directions respectively,
$y_x = d+z-2+\eta$, and $y_\tau = y_x/ z$.
By using these forms, we obtain $y_x=1.44 \pm 0.03$ and $y_\tau=0.92 \pm 0.07$
at $K=0.292$, implying that $z=1.5 \pm 0.1$.
$\xi_x$ grows rapidly as $K \to 0.292$ from below whereas
$\xi_\tau \gg L_\tau$ for $K$ shown in the figures.
The value of $\eta$ is too uncertain to estimate a reliable number.
The error ranges of $y_x$ and $y_\tau$
are estimated as the range of best fittings for different
sizes and aspect ratios. 

This value of the dynamical critical exponent is nearly consistent with
the previously reported value $z=1.35 \pm 0.05$.
Note that we expect $z=2$ for strongly disordered cases
and $z=1$ for weakly disordered cases, if disorder is irrelevant, 
in two dimensions as discussed above.
Therefore, this anomalous value of $z$ suggests that $\Delta_c$
determines the multi-critical line
separating the weak and strong disorder regimes.

What makes these regimes different is an interesting question.
Here we propose that the relevance of
the particle-hole excitations to the transition
is different in these two regimes.
In order to check this conjecture, we plot the density profiles.
The net particle density at ${\bf r}$ on $xy$-plane
for a given disorder complexion is
\be
{\bar n}_{\bf r}=  \langle \frac{1}{L_\tau} \sum_\tau J^\tau_{\bf x} \rangle.
\ee
We define the number of particles and holes as
$N_p=\sum_{\bf r} {\bar n}_{\bf r} \theta({\bar n}_{\bf r})$ and
$N_h=-\sum_{\bf r} {\bar n}_{\bf r} \theta(-{\bar n}_{\bf r})$,
respectively, where $\theta$ is the step function.

\begin{figure}
\includegraphics[width=4.6cm]{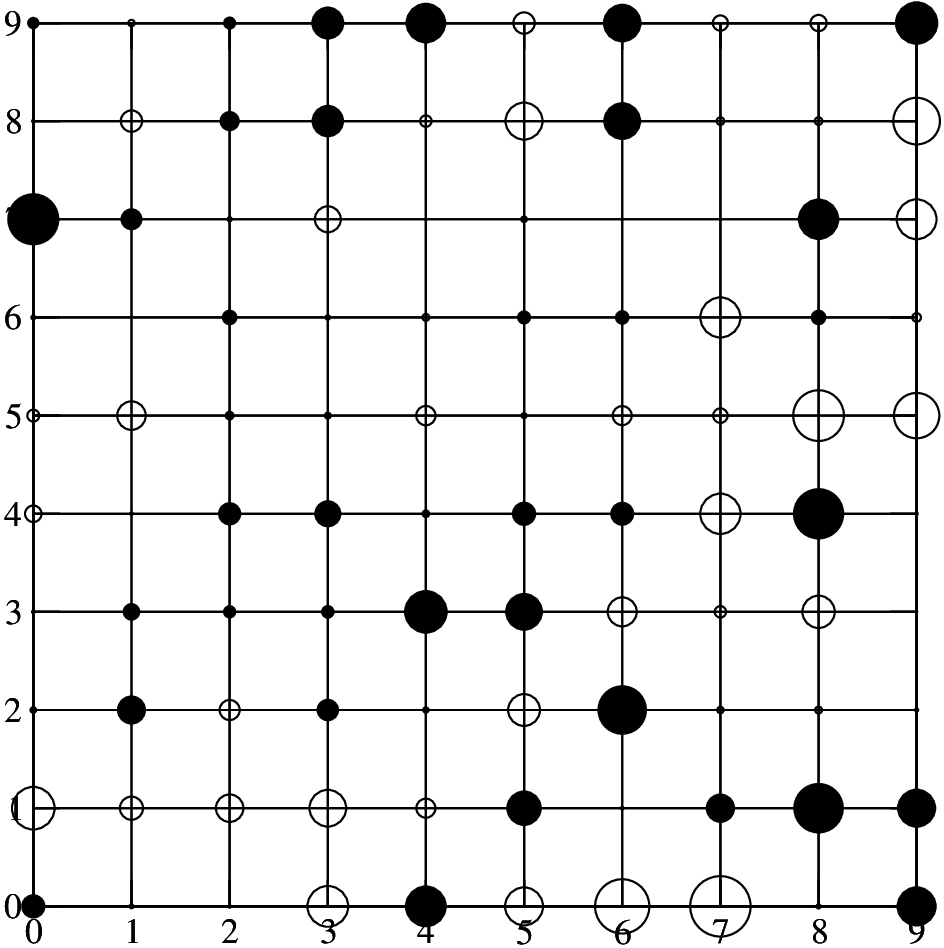}
\caption{
A density plot for commensurate density at $K=0.320$ and $\Delta=0.4$.
Filled (open) circles denote particles (holes). Their density at a certain
site is proportional to the area of the circle centered on that site.
There are particles and holes, and their density grows continuously
as $K$ increases.}
\label{fig:fig3}
\end{figure}

\begin{figure}
\includegraphics[width=4.6cm]{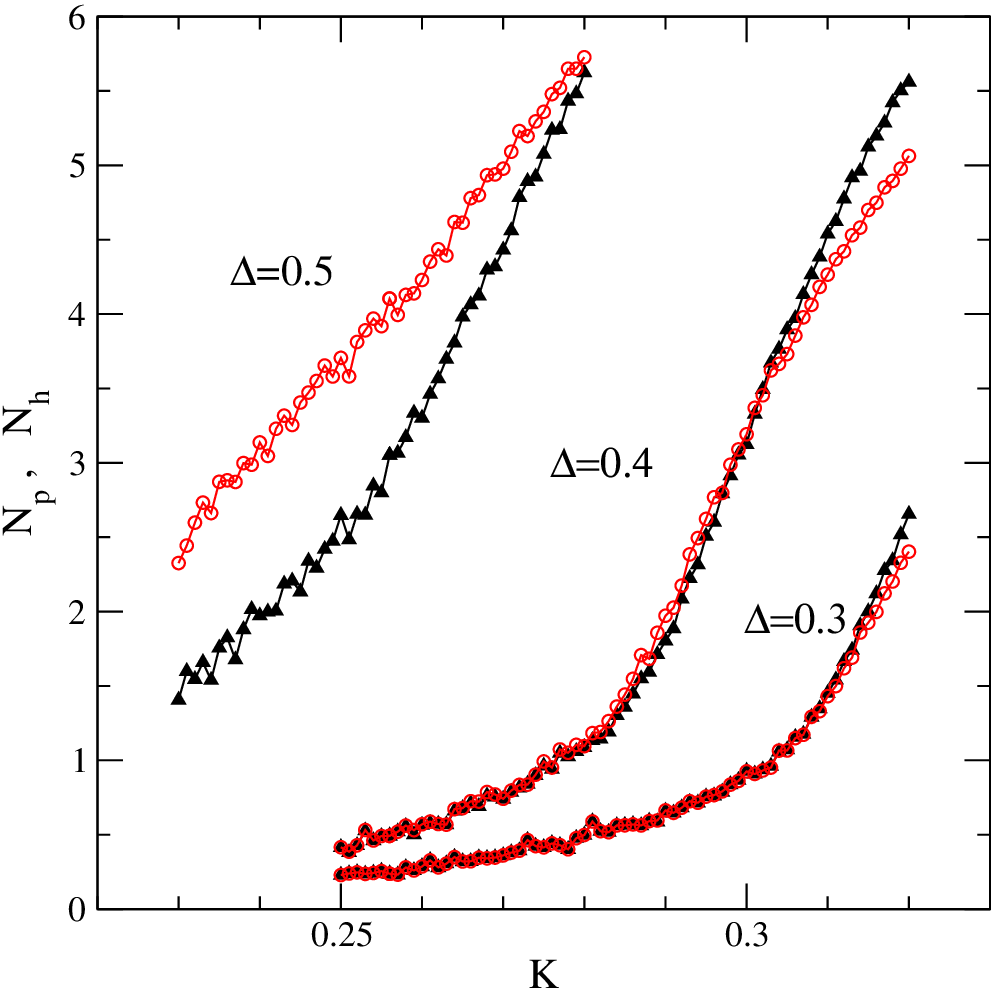}
\caption{
Net number of particles ($N_p$: filled triangles)
and holes ($N_h$: open circles) in a $10 \times 10 \times 25$
lattice as a function of $K$.
The same complexion is used with different $\Delta$.
Lines are guide for eyes.
}
\label{fig:fig4}
\end{figure}

Figure \ref{fig:fig3} is a typical density plot of particles (filled circles)
and holes (open circles) in the presence of disorder for $\Delta=0.4$
at $K=0.320$.
Their density at a certain site is denoted by the area of the circles.
Similar features appear for different $\Delta$ and $K$.
In the Mott insulating regime neither particles nor holes appear.
The fluctuations of particles and holes grow as $K$ increases.
However, they show different behavior in weak and strong disorder regimes,
which is summarized in Figure \ref{fig:fig4}.

Figure \ref{fig:fig4} shows the number of particles and holes
when ${\bar \mu}=0$ for
$\Delta=0.3, 0.4$, and $0.5$ in a $10 \times 10\times 25$ lattice.
The same complexion of disorder is used with different $\Delta$.
Note that the critical points in these systems are
$K_c \approx 0.313, 0.292$, and $0.248$ for 
$\Delta=0.3, 0.4$, and $0.5$, respectively,
and the critical strength of the disorder is $\Delta_c \approx 0.4$.
We see that, for weak disorder ($\Delta=0.3$), the number of particles and holes
matches each other. It strongly suggest that in this regime the low-energy
excitations are the particle-hole pairs.
In addition, their fluctuations increase sharply around the
transition, implying that a finite compressibility rises at the transition
as the energy for density fluctuations vanishes.
All these features supports that a direct MI-to-SF transition actually happens
in this regime.
For strong disorder ($\Delta=0.5$), on the other hand, the numbers of particles
and holes change separately. It means that the single particle or hole states
are occupied separately. The transition occurs as delocalized
single particle states are occupied.
Also the fluctuation increases linearly near the transition.
This implies that there is no abrupt change of the compressibility
at the transition, consistent with the compressible BG insulating picture.
The occurrence of the BG-to-SF transition in this regime has been confirmed
\cite{Wallin94}.
(Note that when $\Delta=0.5$, $K_c$ is independent of $\bar\mu$.)
Therefore, the different behavior of particle and hole fluctuations
as a function of $\Delta$ as shown in Figure \ref{fig:fig4}
supports the existence of the critical strength of disorder
dividing the weak and strong disorder regimes.

In summary,
we calculate the correlation functions of the disordered boson Hubbard model
in the current representation in (2+1) dimensions via Monte Carlo simulations.
For intermediate strength of disorder ($\Delta=0.4$) with commensurate filling
($\bar \mu=0$), we obtain the dynamical critical exponent $z=1.5 \pm 0.1$,
consistent with the finite-size scaling behavior of the superfluid stiffness.
This supports the existence of the multi-critical line separating the strong
and weak disorder regimes.
The number of particles and holes as a function of $K$
strongly supports the scenario that
the density fluctuations due to the particle-hole excitations,
rather than a single-particle excitations,
indeed drive the phase transition in weak disorder regime.
On the contrary, in the strong disorder regime, single particle (hole) states
are relevant to the transition.

\begin{acknowledgements}
This work was supported by Korea Research Foundation Grant
(KRF-2002-041-C00110).
\end{acknowledgements}


\begin{thebibliography}{25}
\expandafter\ifx\csname natexlab\endcsname\relax\def\natexlab#1{#1}\fi
\expandafter\ifx\csname bibnamefont\endcsname\relax
  \def\bibnamefont#1{#1}\fi
\expandafter\ifx\csname bibfnamefont\endcsname\relax
  \def\bibfnamefont#1{#1}\fi
\expandafter\ifx\csname citenamefont\endcsname\relax
  \def\citenamefont#1{#1}\fi
\expandafter\ifx\csname url\endcsname\relax
  \def\url#1{\texttt{#1}}\fi
\expandafter\ifx\csname urlprefix\endcsname\relax\def\urlprefix{URL }\fi
\providecommand{\bibinfo}[2]{#2}
\providecommand{\eprint}[2][]{\url{#2}}

\bibitem[{\citenamefont{Fisher et~al.}(1989)\citenamefont{Fisher, Weichman,
  Grinstein, and Fisher}}]{Fisher89}
\bibinfo{author}{\bibfnamefont{M.~P.~A.} \bibnamefont{Fisher}},
  \bibinfo{author}{\bibfnamefont{P.~B.} \bibnamefont{Weichman}},
  \bibinfo{author}{\bibfnamefont{G.}~\bibnamefont{Grinstein}},
  \bibnamefont{and} \bibinfo{author}{\bibfnamefont{D.~S.}
  \bibnamefont{Fisher}}, \bibinfo{journal}{Phys. Rev. B}
  \textbf{\bibinfo{volume}{40}}, \bibinfo{pages}{546} (\bibinfo{year}{1989}).

\bibitem[{\citenamefont{Scalettar et~al.}(1991)\citenamefont{Scalettar,
  Batrouni, and Zim\'anyi}}]{Scalettar91}
\bibinfo{author}{\bibfnamefont{R.~T.} \bibnamefont{Scalettar}},
  \bibinfo{author}{\bibfnamefont{G.~G.} \bibnamefont{Batrouni}},
  \bibnamefont{and} \bibinfo{author}{\bibfnamefont{G.~T.}
  \bibnamefont{Zim\'anyi}}, \bibinfo{journal}{Phys. Rev. Lett.}
  \textbf{\bibinfo{volume}{66}}, \bibinfo{pages}{3144} (\bibinfo{year}{1991}).

\bibitem[{\citenamefont{Krauth and Trivedi}(1991)}]{Krauth91b}
\bibinfo{author}{\bibfnamefont{W.}~\bibnamefont{Krauth}} \bibnamefont{and}
  \bibinfo{author}{\bibfnamefont{N.}~\bibnamefont{Trivedi}},
  \bibinfo{journal}{Europhys. Lett.} \textbf{\bibinfo{volume}{14}},
  \bibinfo{pages}{627} (\bibinfo{year}{1991}).

\bibitem[{\citenamefont{Singh and Rokhsar}(1992)}]{Singh92}
\bibinfo{author}{\bibfnamefont{K.~G.} \bibnamefont{Singh}} \bibnamefont{and}
  \bibinfo{author}{\bibfnamefont{D.~S.} \bibnamefont{Rokhsar}},
  \bibinfo{journal}{Phys.Rev. B} \textbf{\bibinfo{volume}{46}},
  \bibinfo{pages}{3002} (\bibinfo{year}{1992}).

\bibitem[{\citenamefont{Runge}(1992)}]{Runge92}
\bibinfo{author}{\bibfnamefont{K.~J.} \bibnamefont{Runge}},
  \bibinfo{journal}{Phys. Rev. B} \textbf{\bibinfo{volume}{43}},
  \bibinfo{pages}{13136} (\bibinfo{year}{1992}).

\bibitem[{\citenamefont{P\'azm\'andi and Zim\'anyi}(1998)}]{Pazmandi98}
\bibinfo{author}{\bibfnamefont{F.}~\bibnamefont{P\'azm\'andi}}
  \bibnamefont{and} \bibinfo{author}{\bibfnamefont{G.~T.}
  \bibnamefont{Zim\'anyi}}, \bibinfo{journal}{Phys. Rev. B}
  \textbf{\bibinfo{volume}{57}}, \bibinfo{pages}{5044} (\bibinfo{year}{1998}).

\bibitem[{\citenamefont{Zhang et~al.}(1995)\citenamefont{Zhang, Kawashima,
  Carlson, and Gubernatis}}]{Zhang95}
\bibinfo{author}{\bibfnamefont{S.}~\bibnamefont{Zhang}},
  \bibinfo{author}{\bibfnamefont{N.}~\bibnamefont{Kawashima}},
  \bibinfo{author}{\bibfnamefont{J.}~\bibnamefont{Carlson}}, \bibnamefont{and}
  \bibinfo{author}{\bibfnamefont{J.~E.} \bibnamefont{Gubernatis}},
  \bibinfo{journal}{Phys. Rev. Lett.} \textbf{\bibinfo{volume}{74}},
  \bibinfo{pages}{1500} (\bibinfo{year}{1995}).

\bibitem[{\citenamefont{Amico and Penna}(1998)}]{Amico98}
\bibinfo{author}{\bibfnamefont{L.}~\bibnamefont{Amico}} \bibnamefont{and}
  \bibinfo{author}{\bibfnamefont{V.}~\bibnamefont{Penna}},
  \bibinfo{journal}{Phys. Rev. Lett.} \textbf{\bibinfo{volume}{80}},
  \bibinfo{pages}{2189} (\bibinfo{year}{1998}).

\bibitem[{\citenamefont{Herbut}(2000)}]{Herbut00}
\bibinfo{author}{\bibfnamefont{I.~F.} \bibnamefont{Herbut}},
  \bibinfo{journal}{Phys. Rev. B} \textbf{\bibinfo{volume}{61}},
  \bibinfo{pages}{14723} (\bibinfo{year}{2000}).

\bibitem[{\citenamefont{Wallin et~al.}(1994)\citenamefont{Wallin, S{\o}rensen,
  Girvin, and Young}}]{Wallin94}
\bibinfo{author}{\bibfnamefont{M.}~\bibnamefont{Wallin}},
  \bibinfo{author}{\bibfnamefont{E.~S.} \bibnamefont{S{\o}rensen}},
  \bibinfo{author}{\bibfnamefont{S.~M.} \bibnamefont{Girvin}},
  \bibnamefont{and} \bibinfo{author}{\bibfnamefont{A.~P.} \bibnamefont{Young}},
  \bibinfo{journal}{Phys. Rev. B} \textbf{\bibinfo{volume}{49}},
  \bibinfo{pages}{12115} (\bibinfo{year}{1994}).

\bibitem[{\citenamefont{Crowell et~al.}(1997)\citenamefont{Crowell, Keuls, and
  Reppy}}]{Crowell97}
\bibinfo{author}{\bibfnamefont{P.~A.} \bibnamefont{Crowell}},
  \bibinfo{author}{\bibfnamefont{F.~W.~V.} \bibnamefont{Keuls}},
  \bibnamefont{and} \bibinfo{author}{\bibfnamefont{J.~D.} \bibnamefont{Reppy}},
  \bibinfo{journal}{Phys. Rev. B} \textbf{\bibinfo{volume}{55}},
  \bibinfo{pages}{12620} (\bibinfo{year}{1997}).

\bibitem[{\citenamefont{Greiner et~al.}(2002)\citenamefont{Greiner, Mandel,
  Esslinger, H\"ansch, and Bloch}}]{Greiner02}
\bibinfo{author}{\bibfnamefont{M.}~\bibnamefont{Greiner}},
  \bibinfo{author}{\bibfnamefont{O.}~\bibnamefont{Mandel}},
  \bibinfo{author}{\bibfnamefont{T.}~\bibnamefont{Esslinger}},
  \bibinfo{author}{\bibfnamefont{T.}~\bibnamefont{H\"ansch}}, \bibnamefont{and}
  \bibinfo{author}{\bibfnamefont{I.}~\bibnamefont{Bloch}},
  \bibinfo{journal}{Nature (London)} \textbf{\bibinfo{volume}{415}},
  \bibinfo{pages}{39} (\bibinfo{year}{2002}).

\bibitem[{\citenamefont{Damski et~al.}(2003)\citenamefont{Damski, Zakrzewski,
  Santos, Zoller, and Lewenstein}}]{Zoller03}
\bibinfo{author}{\bibfnamefont{B.}~\bibnamefont{Damski}},
  \bibinfo{author}{\bibfnamefont{J.}~\bibnamefont{Zakrzewski}},
  \bibinfo{author}{\bibfnamefont{L.}~\bibnamefont{Santos}},
  \bibinfo{author}{\bibfnamefont{P.}~\bibnamefont{Zoller}}, \bibnamefont{and}
  \bibinfo{author}{\bibfnamefont{M.}~\bibnamefont{Lewenstein}},
  \bibinfo{journal}{Phys. Rev. Lett.} \textbf{\bibinfo{volume}{91}},
  \bibinfo{pages}{080403} (\bibinfo{year}{2003}).

\bibitem[{\citenamefont{Goldman and Markovi\'c}(1998)}]{Goldman98}
\bibinfo{author}{\bibfnamefont{A.~M.} \bibnamefont{Goldman}} \bibnamefont{and}
  \bibinfo{author}{\bibfnamefont{N.}~\bibnamefont{Markovi\'c}},
  \bibinfo{journal}{Physics Today} \textbf{\bibinfo{volume}{51}},
  \bibinfo{pages}{No. 11 39} (\bibinfo{year}{1998}).

\bibitem[{\citenamefont{Fazio and van~der Zant}(2001)}]{Fazio01}
\bibinfo{author}{\bibfnamefont{R.}~\bibnamefont{Fazio}} \bibnamefont{and}
  \bibinfo{author}{\bibfnamefont{H.}~\bibnamefont{van~der Zant}},
  \bibinfo{journal}{Phys. Rep.} \textbf{\bibinfo{volume}{355}},
  \bibinfo{pages}{235} (\bibinfo{year}{2001}).

\bibitem[{\citenamefont{Gerber et~al.}(1997)\citenamefont{Gerber, Milner,
  Deutscher, Karpovsky, and Gladkikh}}]{Gerber97}
\bibinfo{author}{\bibfnamefont{A.}~\bibnamefont{Gerber}},
  \bibinfo{author}{\bibfnamefont{A.}~\bibnamefont{Milner}},
  \bibinfo{author}{\bibfnamefont{G.}~\bibnamefont{Deutscher}},
  \bibinfo{author}{\bibfnamefont{M.}~\bibnamefont{Karpovsky}},
  \bibnamefont{and} \bibinfo{author}{\bibfnamefont{A.}~\bibnamefont{Gladkikh}},
  \bibinfo{journal}{Phys. Rev. Lett.} \textbf{\bibinfo{volume}{78}},
  \bibinfo{pages}{4277} (\bibinfo{year}{1997}).

\bibitem[{\citenamefont{Prokof'ev and Svitsunov}(2003)}]{Prokofev03}
\bibinfo{author}{\bibfnamefont{P.}~\bibnamefont{Prokof'ev}} \bibnamefont{and}
  \bibinfo{author}{\bibfnamefont{B.}~\bibnamefont{Svitsunov}},
  \bibinfo{journal}{cond-mat/0301205 (unpublished)}  (\bibinfo{year}{2003}).

\bibitem[{\citenamefont{Phillips and Dalidovich}(2003)}]{Phillips03}
\bibinfo{author}{\bibfnamefont{P.}~\bibnamefont{Phillips}} \bibnamefont{and}
  \bibinfo{author}{\bibfnamefont{D.}~\bibnamefont{Dalidovich}},
  \bibinfo{journal}{Science} \textbf{\bibinfo{volume}{302}},
  \bibinfo{pages}{243} (\bibinfo{year}{2003}).

\bibitem[{\citenamefont{Freericks and Monien}(1996)}]{Freericks96}
\bibinfo{author}{\bibfnamefont{J.~K.} \bibnamefont{Freericks}}
  \bibnamefont{and} \bibinfo{author}{\bibfnamefont{H.}~\bibnamefont{Monien}},
  \bibinfo{journal}{Phys.Rev. B} \textbf{\bibinfo{volume}{53}},
  \bibinfo{pages}{2691} (\bibinfo{year}{1996}).

\bibitem[{\citenamefont{Mukhopadhyay and Weichman}(1996)}]{Weichman96}
\bibinfo{author}{\bibfnamefont{R.}~\bibnamefont{Mukhopadhyay}}
  \bibnamefont{and} \bibinfo{author}{\bibfnamefont{P.}~\bibnamefont{Weichman}},
  \bibinfo{journal}{Phys. Rev. Lett.} \textbf{\bibinfo{volume}{76}},
  \bibinfo{pages}{2977} (\bibinfo{year}{1996}).

\bibitem[{\citenamefont{Kisker and Rieger}(1997)}]{Kisker97}
\bibinfo{author}{\bibfnamefont{J.}~\bibnamefont{Kisker}} \bibnamefont{and}
  \bibinfo{author}{\bibfnamefont{H.}~\bibnamefont{Rieger}},
  \bibinfo{journal}{Phys. Rev. B} \textbf{\bibinfo{volume}{55}},
  \bibinfo{pages}{R11981} (\bibinfo{year}{1997}).

\bibitem[{\citenamefont{Park et~al.}(1999)\citenamefont{Park, Lee, Cha, Choi,
  Kim, and Kim}}]{Park99}
\bibinfo{author}{\bibfnamefont{S.~Y.} \bibnamefont{Park}},
  \bibinfo{author}{\bibfnamefont{J.-W.} \bibnamefont{Lee}},
  \bibinfo{author}{\bibfnamefont{M.-C.} \bibnamefont{Cha}},
  \bibinfo{author}{\bibfnamefont{M.~Y.} \bibnamefont{Choi}},
  \bibinfo{author}{\bibfnamefont{B.~J.} \bibnamefont{Kim}}, \bibnamefont{and}
  \bibinfo{author}{\bibfnamefont{D.}~\bibnamefont{Kim}},
  \bibinfo{journal}{Phys. Rev. B} \textbf{\bibinfo{volume}{59}},
  \bibinfo{pages}{8420} (\bibinfo{year}{1999}).

\bibitem[{\citenamefont{Lee et~al.}(2001)\citenamefont{Lee, Cha, and
  Kim}}]{Lee01}
\bibinfo{author}{\bibfnamefont{J.-W.} \bibnamefont{Lee}},
  \bibinfo{author}{\bibfnamefont{M.-C.} \bibnamefont{Cha}}, \bibnamefont{and}
  \bibinfo{author}{\bibfnamefont{D.}~\bibnamefont{Kim}},
  \bibinfo{journal}{Phys.\ Rev. Lett.} \textbf{\bibinfo{volume}{87}},
  \bibinfo{pages}{247006} (\bibinfo{year}{2001}).

\bibitem[{\citenamefont{Alet and S{\o}rensen}(2003{\natexlab{a}})}]{Alet03a}
\bibinfo{author}{\bibfnamefont{F.}~\bibnamefont{Alet}} \bibnamefont{and}
  \bibinfo{author}{\bibfnamefont{E.~S.} \bibnamefont{S{\o}rensen}},
  \bibinfo{journal}{Phys. Rev. E} \textbf{\bibinfo{volume}{67}},
  \bibinfo{pages}{015701(R)} (\bibinfo{year}{2003}{\natexlab{a}}).

\bibitem[{\citenamefont{Alet and S{\o}rensen}(2003{\natexlab{b}})}]{Alet03b}
\bibinfo{author}{\bibfnamefont{F.}~\bibnamefont{Alet}} \bibnamefont{and}
  \bibinfo{author}{\bibfnamefont{E.~S.} \bibnamefont{S{\o}rensen}},
  \bibinfo{journal}{Phys. Rev. E} \textbf{\bibinfo{volume}{68}},
  \bibinfo{pages}{026702} (\bibinfo{year}{2003}{\natexlab{b}}).

\end{thebibliography}
\end{document}